\documentclass{ws-mpla}

\begin{document}

\markboth{Juan M. Romero, J. A. Santiago, O. Gonz\'alez-Gaxiola, A. Zamora }
{ Electrodynamics \`a la Ho\v{r}ava  }

\catchline{}{}{}{}{}

\title{ Electrodynamics \`a la Ho\v{r}ava }

\author{\footnotesize Juan M. Romero$^{\dagger}$, Jos\'e~A.~Santiago $^{\ddagger}$,
 O. Gonz\'alez-Gaxiola$^{\ast}$, Adolfo Zamora$^{\heartsuit}$  }
\address{
Departamento de Matem\'aticas Aplicadas y Sistemas,\\  
Universidad Aut\'onoma Metropolitana-Cuajimalpa,\\
M\'exico, D.F  01120, M\'exico\\
$\dagger$jromero@correo.cua.uam.mx,\\
$\ddagger$jsantiago@correo.cua.uam.mx,\\
$\ast$ ogonzalez@correo.cua.uam.mx,\\
$\heartsuit$ zamora@correo.cua.uam.mx}

\maketitle

\pub{Received (Day Month Year)}{Revised (Day Month Year)}

\begin{abstract}
We study an electrodynamics consistent with anisotropic transformations
of space-time with an arbitrary dynamic exponent $z$. The equations of
motion and conserved quantities are explicitly obtained. We  show that the propagator 
of this theory can be regarded as a quantum correction
to the usual propagator. Moreover we obtain that both the momentum and angular momentum are not modified, but 
their conservation laws do change. We also show that in this theory the
speed of light and the electric charge are modified with $z$. 
The magnetic monopole in this electrodynamics and its duality 
transformations are also investigated. For that we found that there exists 
a dual electrodynamics, with higher derivatives in the electric field, 
invariant under the same anisotropic transformations.
\end{abstract}

\ccode{PACS Nos.: 11.10.-z,  11.30.-j}

\section{Introduction}
\label{s:Intro}
In several areas of physics the scale-invariant systems are important.
These systems have been recognized in the context of condensed matter
\cite{condensada:gnus,cardy:gnus}
and theories of modified gravity such as MOND \cite{milgrom:gnus},
see also \cite{mond-horava:gnus}. Of particular importance are the
systems invariant under anisotropic scale-transformations of the
space-time of the form
\begin{equation}
t\to b^{z}t,\qquad \vec x\to b \vec x,
\label{eq:escala}
\end{equation}
where $z$ plays the role of a dynamical critical-exponent. These 
transformations arise in the non-relativistic limit of string theory
\cite{string-1:gnus,string-2:gnus} and, through the duality $AdS/CFT$, this theory can
be related to condensed matter systems \cite{herzog1:gnus,herzog2:gnus}.
Recently, in addition, P.~Ho\v{r}ava proposed a gravity compatible with (\ref{eq:escala}). This gravity is non-relativistic
but at large distances retrieves a gravity similar to Einstein's
\cite{horava:gnus}. A noteworthy fact of Ho\v{r}ava gravity is that
replaces the usual dispersion relation by
\begin{equation}
k_{0}^{2}-G( k^{2})^{z}=0, \qquad G={\rm constant}, k^{2}=k_{1}^{2}+k_{2}^{2}+k^{2}_{3},
\label{eq:Horava}
\end{equation}
but more importantly, for the case of $z=3$, yields a ghost-free gravity 
which is renormalizable by power counting. This new gravity has features
that makes it notable; for instance, in a natural way has an alternative
mechanism to inflation \cite{brandenberger:gnus}. In addition, it can
explain some cosmological phenomena without introducing dark matter
\cite{mukohyama:gnus}; some systems in this gravity can be found in
\cite{bibhas-1:gnus,bibhas-2:gnus}. Another feature is that the dispersion relation
(\ref{eq:Horava}) arises not from a geodesic equation but from a mechanics 
invariant under (\ref{eq:escala}), 
see \cite{romero1:gnus,romero2:gnus,romero3:gnus,romero4:gnus,romero5:gnus,romero6:gnus}. 
Ho\v{r}ava gravity is recent and has several aspects not well understood; for 
example, it presents some dynamical problems \cite{pag-1:gnus,pag-2:gnus,pag-3:gnus,pag-4:gnus,pag-5:gnus,pag-6:gnus}; see
also \cite{venezuela:gnus}. A modification to Ho\v{r}ava gravity, free 
from these problems, can be found in \cite{blas:gnus,blas-1:gnus,blas-2:gnus}; however see 
\cite{tony:gnus}. From all this, Ho\v{r}ava gravity is without a doubt 
an important step towards understanding the quantum aspects of gravity.\\

Field theories consistent with the dispersion relation (\ref{eq:Horava})
for the case of $z=2$ have been studied \cite{das:gnus,andreev:gnus,nakayama:gnus,pavlopoulos:gnus}, 
some systems with $z=3$ can be seen in 
\cite{tpsalis-1:gnus,tpsalis-2:gnus}. Systems with arbitrary $z$ have been studied in: black holes
\cite{amanda:gnus}, nonrelativistic $AdS/CFT$ duality \cite{hartnoll:gnus,malda:gnus}, and the causal structure of Schr\"odinger space
\cite{blau:gnus}.
 However, the case of $z$ arbitrary has been seldom
considered in field theory. In this paper, based on the initial works \cite{fradkin:gnus,horava3:gnus} 
and \cite{chen:gnus}, we study an electrodynamics compatible with the
transformations (\ref{eq:escala}) for arbitrary $z$. Modified Maxwell's 
equations and their conserved quantities are first obtained for the
case of $z=2$ and then, based on these results, a general model that includes derivatives of higher 
order in the magnetic field is constructed. For this model we obtain
a dispersion relation of the form
\begin{equation}
\frac{\omega^{2}\alpha}{c^{2}}=\sum_{n\geq 1}b_{n}\left(k^{2}\right)^{n}, 
\label{eq:Horava-disper-0}
\end{equation}
and show that $\alpha$ acts as a refraction index. 
Notice that the dispersion relation (\ref{eq:Horava}) is a particular case of (\ref{eq:Horava-disper-0}). The conserved
quantities are also obtained; in particular it is shown that energy is
modified, but the momentum and angular momentum remain unchanged. In
addition, we show the existence of a conserved quantity additional to
the usual ones. For this system we propose the modified Maxwell's equations
with sources and obtain their conservation laws. In particular, the
modifications to the stress tensor are provided.  Moreover, from duality
transformations we consider the problem of the magnetic monopole and
show the existence of a dual electrodynamics compatible with the
dispersion relation (\ref{eq:Horava-disper-0}). Finally, we  show that the propagator 
of this theory  can be regarded  as a quantum correction to
the usual Maxwell's propagator. \\

It is worth mentioning that non-relativistic Maxwell theory  has been
already studied in another context \cite{hussin:gnus,yifu:gnus,nakayam:gnus}. Other 
models with higher derivatives in the magnetic fields can be found in
\cite{kiritsis:gnus,maeda:gnus,gruss:gnus}.\\

This paper is organized as follows: in section \ref{sn:El-H-z-2} 
Ho\v{r}ava electrodynamics for the case of $z=2$ is studied; in the section
\ref{sn:El-H-z-gen}  we present the general case of arbitrary $z$; an analysis
of the magnetic monopole question appears in section \ref{sn:dual};  we obtain  
the  Green function in section 5;  finally our results are summarized in section \ref{sn:summ}.

\section{Electrodynamics \`a la Ho\v{r}ava: $z=2$}
\label{sn:El-H-z-2}

In this section we study Ho\v{r}ava electrodynamics for the case of $z=2$
originally proposed in \cite{horava3:gnus}. For completeness we recall the
definitions
\begin{eqnarray}
E_{i}=-\left(\partial_{ct}A_{i}+\partial_{i}\phi\right),\quad 
B_{i}= \left(\vec\nabla \times \vec A\right)_{i},\quad 
F_{ij}=\partial_{i}A_{j}-\partial_{j}A_{i}=\epsilon_{ijk}B_{k}.\label{eq:def3}
\end{eqnarray}
Now, let us assume we have the action
\begin{equation}
S=\int cdt d\vec x {\cal  L}= \int cdt d\vec x \left(\alpha E_{i}E_{i} 
+\beta \partial_{i}F_{ik}\partial_{j}F_{jk}\right).
\end{equation}
By requesting $\delta S=0$ one finds the equations of motion
\begin{eqnarray}
\alpha\partial_{ct}E_{k} + \beta \partial^{2}\partial_{i}F_{ik}=0,\qquad 
\partial_{i}E_{i}&=&0.
\end{eqnarray}
Consideration of (\ref{eq:def3}) leads to the modified 
Maxwell's equations
\begin{eqnarray}
\vec\nabla\cdot \vec E&=& 0\label{eq:m-h1},\\
\vec\nabla\times \vec E&=&-\frac{1}{c}\frac{\partial \vec B}{\partial t}
\label{eq:m-h2},\\
\vec \nabla\cdot \vec B&=&0,\label{eq:m-h3}\\
\vec\nabla\times\left( \beta \nabla^{2}\vec B\right)&=&
\frac{\alpha}{c}\frac{\partial \vec E}{\partial t}.\label{eq:m-h4}
\end{eqnarray}
From (\ref{eq:m-h2}) and (\ref{eq:m-h4}) we obtain
\begin{eqnarray}
\left(\beta \left(\nabla^{2}\right)^{2}-\frac{\alpha}{c^{2}}
\frac{\partial^{2}}{\partial t^{2}}\right)\vec E=0,\qquad 
\left(\beta \left(  \nabla^{2}\right)^{2}-\frac{\alpha}{c^{2}}
\frac{\partial^{2}}{\partial t^{2}}\right)\vec B&=&0.
\end{eqnarray}
Then, the dispersion relation for plane waves is
$\beta \left(k^{2}\right)^{2}=-{\alpha \omega^{2}}/{c^{2}}$.\\

The equations (\ref{eq:m-h1})-(\ref{eq:m-h4}) are invariant under the transformations of scale
\begin{equation}
\vec E \to b^{-3} \vec E, \quad\! \vec B \to b^{-2} \vec B,\quad\! 
t\to b^{2}t,\quad\! \vec x \to b \vec x, \label{eq:escalz2}
\end{equation}
however, the action transforms as
\begin{eqnarray}
S\to b^{-1}S.
\end{eqnarray}
A similar case occurs to the harmonic oscillator, whose equations of
motion are invariant under the scale transformations $x\to bx$, but the 
action does not have this invariance.\\

\subsection{Conserved quantities} 
\label{ssn:cant-cons}

By considering the equations of motion (\ref{eq:m-h1})-(\ref{eq:m-h4}) one may show that the quantities
\begin{eqnarray}
{\cal E}&=&\frac{c}{8\pi}\int d\vec x \left( \alpha\vec E^{2}
+\beta \vec B\cdot \nabla^{2} \vec B\right),\\
\vec P&=&\frac{1}{4\pi c}\int d\vec x \left(\alpha \vec E\times 
\vec B\right),\\
\vec L &=&\frac{1}{4\pi c} \int d\vec x \left[\vec x \times 
\left(\alpha \vec E\times \vec B\right)\right],
\end{eqnarray}
are conserved. From these we see that the energy is different to the 
usual one, but both the momentum and angular momentum remain unchanged.
Below  we will find a generalization of these quantities and their 
conservation laws for arbitrary $z$.\\

Although Noether's theorem does not give a conserved charge for symmetry (\ref{eq:escalz2}), 
in the next section we will  associate
a conserved charge with this.

\section{Electrodynamics \`a la Ho\v{r}ava: General Case}
\label{sn:El-H-z-gen}

We now move on to propose a model that allows one to obtain a dispersion
relation of the form (\ref{eq:Horava}) valid for arbitrary $z$. By
assuming $f(x)$ a smooth function, we pose the action
\begin{eqnarray}
S&=& \int cdt d\vec x {\cal  L}= \int cdt d\vec x \left(\alpha E_{i}E_{i} 
-\vec B \cdot f\left(\nabla ^{2}\right)\vec B\right)\nonumber\\
&=&  \int cdt d\vec x \left(\alpha E_{i}E_{i} -\frac{1}{2} F_{ij} 
f\left(\nabla ^{2}\right)F_{ij}\right).
\end{eqnarray}
First let us observe that in units of momenta we have
\begin{eqnarray}
[cdt dx^{3}]=-(3+z),\quad [E_{i}]=(z+1),\quad [B_{i}]=2,\quad 
\left[ \alpha \right] =  -( z-1 ).
\end{eqnarray}
Now, if $f(x)=\sum_{n\geq 1} a_{n}x^{n-1}$ and $[a_{n}]=z-1-2(n-1)$, then $
\left[f\left(\nabla ^{2}\right)\right]=z-1;$ therefore, $S$ has no units.\\

Note us that the Lagrangian can be regarded  as
\begin{eqnarray}
{\cal L}={\cal L}_{0}+{\cal L}_{I},
\end{eqnarray}
where, 
\begin{eqnarray}
{\cal L}_{0}=\alpha \vec E\cdot \vec E-a_{1} \vec B \cdot \vec B
\end{eqnarray}
the free part of the Lagrangian and 
\begin{eqnarray}
{\cal L}_{I}=\sum_{n=2}^{\infty}a_{n}\vec B\cdot \left(\nabla^{2}\right)^{n-1}\vec B,
\end{eqnarray}
the interaction one. \\

Before performing the variation of $S$, notice that carrying out
$n$ integrations by parts and neglecting the boundary terms we obtain
\begin{equation}
\int d\vec x F_{ij}\left(\nabla^{2}\right)^{n}\delta F_{ij}=
(-)^{2n} \int d\vec x \delta F_{ij}\left(\nabla^{2}\right)^{n}F_{ij},
\end{equation}
so that
\begin{eqnarray}
 \int d\vec x \delta \left(F_{ij} \left(\nabla^{2}\right)^{n} F_{ij}\right)
&=&\int d\vec x \left( \delta F_{ij}\left(\nabla^{2}\right)^{n}F_{ij}+ F_{ij}
\left(\nabla^{2}\right)^{n}\delta F_{ij} \right)\nonumber\\
&=& 
2\int d\vec x \delta F_{ij}\left(\nabla^{2}\right)^{n}F_{ij},
\end{eqnarray}
from where we may define
\begin{equation}
\int d\vec x  F_{ij}f\left(\nabla^{2}\right) F_{ij}.
\end{equation}
Thus
\begin{eqnarray}
& &\delta \int d\vec x F_{ij}f\left(\nabla^{2}\right) F_{ij}
=\int d\vec x \delta \left( F_{ij}f\left(\nabla^{2}\right)\right)F_{ij}\nonumber\\
&=&\int d\vec x \delta \left( F_{ij}\left[ \sum_{n\geq 1} a_{n}
\left(\nabla^{2}\right)^{n-1} \right]F_{ij}\right)
=\sum_{n\geq 1} a_{n} \int d\vec x \delta \left( 
F_{ij}\left(\nabla^{2}\right)^{n-1} F_{ij}\right)\nonumber\\
&=&2 \sum_{n\geq 1} a_{n} \int d\vec x \delta F_{ij}\left(
\nabla^{2}\right)^{n-1}F_{ij}=2 \int d\vec x \delta F_{ij}f\left(\nabla^{2}\right)F_{ij}.
\end{eqnarray}
By using definition (\ref{eq:def3}) and neglecting  boundary terms one
gets to
\begin{equation}
\delta \int d\vec x F_{ij}f\left(\nabla^{2}\right) F_{ij}
=-4 \int d\vec x \delta A_{j}\partial_{i}
\left( f\left(\nabla^{2}\right)F_{ij}\right).
\end{equation}
Taking these results into account and once again neglecting  boundary terms
one obtains
\begin{equation}
\delta S=\int cdt d\vec x  2\Big[\delta A_{i}\big(f(\nabla^{2})
\partial_{j}F_{ji}+\alpha\partial_{ct}E_{i}\big)
+\alpha \delta\phi \partial_{i}E_{i}\Big]. 
\end{equation}
Therefore, $\delta S=0$ implies the equations of motion
\begin{eqnarray}
\vec\nabla\cdot \vec E&=&0,\label{eq:m-general-1}\\
\vec \nabla \times \vec E&=&-\frac{1}{c}\frac{\partial \vec B}{\partial t}, 
\label{eq:m-general-2}\\
\vec\nabla\cdot \vec B&=&0,\label{eq:m-general-3}\\
\vec\nabla\times\left( f\left(\nabla^{2}\right)\vec B\right)&=&
\frac{\alpha }{c}\frac{\partial \vec E}{\partial t}.\label{eq:m-general-4}
\end{eqnarray}
For $f(x)=a_{z}x^{z-1}$ these yield
\begin{eqnarray}
\vec\nabla\cdot \vec E&=&0,\qquad 
\vec \nabla \times \vec E=-\frac{1}{c}\frac{\partial \vec B}{\partial t}, \label{eq:escala-ecuaciones1}\\
\vec\nabla\cdot \vec B&=&0,\qquad
\vec\nabla\times\left( a_{z}\left(\nabla^{2}\right)^{z-1}\vec B\right)=
\frac{\alpha }{c}\frac{\partial \vec E}{\partial t}.\label{eq:escala-ecuaciones2}
\end{eqnarray}
Notice  that the dimensions of $\alpha$ change for each $z$. As $[\alpha]=-(z-1)$, we may assume that $\alpha$ is
of the form $l^{z-1}$, where $l$ is a constant with units of length. Under
this assumption, it is clear that $\alpha=1$ for $z=1$, but $\alpha \not =1$ 
for $z\not =1$. Therefore, the constant $\alpha$ changes with $z.$\\

In addition to Eqs. (\ref{eq:m-general-2}) and (\ref{eq:m-general-4}) we
obtain the modified wave equations
\begin{eqnarray}
\left(f\left(\nabla^{2}\right)\nabla^{2}-\frac{\alpha}{c^{2}}
\frac{\partial^{2}}{\partial t^{2}}\right)\vec E&=&0,\label{eq:onda-1}\\
\left(f\left(\nabla^{2}\right) \nabla^{2}-\frac{\alpha}{c^{2}}
\frac{\partial^{2}}{\partial t^{2}}\right)\vec B&=&0.\label{eq:onda-2}
\end{eqnarray}
For the case of plane waves one obtains the dispersion relation
\begin{equation}
\frac{\omega^{2} \alpha}{c^{2}}=f(-k^{2})k^{2}=\sum_{n\geq 1}(-)^{n-1}a_{n}\left(k^{2}\right)^{n}. 
\label{eq:dispersion-general}
\end{equation}
In particular, if $a_{n}=0$ for $n\not =z$ and $a_{z}=(-)^{z-1}G$, this yield
\begin{equation}
\frac{\omega^{2} \alpha}{c^{2}}= G(k^{2})^{z},
\end{equation}
which is Ho\v{r}ava dispersion relation (\ref{eq:Horava}). By defining 
$c^{\prime}={c}/{n}$ with $n=\sqrt{\alpha}$, we can consider $\alpha$ as a
refraction index that changes with $z.$

\subsection{Case with sources}
\label{ssn:fuentes}

The modified Maxwell's equations with sources are
\begin{eqnarray}
\vec\nabla\cdot \vec E&=&4\pi \rho \label{eq:m1},\\
\vec\nabla\times \vec E&=&-\frac{1}{c}\frac{\partial \vec B}{\partial t}
\label{eq:m2},\\
\vec \nabla\cdot \vec B&=&0,\label{eq:m3}\\
\vec\nabla\times\left( f\left(\nabla^{2}\right)\vec B\right)&=&\frac{4\pi}{c}
\vec J+\frac{\alpha}{c}\frac{\partial \vec E}{\partial t}.\label{eq:m4}
\end{eqnarray}
By considering equations (\ref{eq:m1}) and (\ref{eq:m4}) we obtain conservation
of the electric charge:
\begin{equation}
\frac{\partial \rho^{\prime}}{\partial t}+\vec \nabla \cdot \vec J=0,
\end{equation}
with $\rho^{\prime}=\alpha\rho$. This implies that the electric charge is
modified by  $z$. For instance, for a point particle the electric charge
changes from $e$ to $e^{\prime}=\alpha e$. Below  we will find this 
phenomenon also present in the Lorentz force.\\

\subsection{Conserved quantities}
\label{ssn:conservadas}

For $\rho=0$ and $\vec J=\vec 0$ one obtains the conserved quantities
\begin{eqnarray}
{\cal E}&=&\frac{c}{8\pi}\int d\vec x \left(\alpha\vec E\cdot\vec E
+\vec B\cdot f\left(\nabla^{2}\right) \vec B\right),\label{eq:conservadas1}\\
\vec P&=&\frac{1}{4\pi c}\int d\vec x\left(\alpha\vec E\times\vec B\right),\\
\vec L &=&\frac{1}{4\pi c}\int d\vec x \left[\vec x \times \left(\alpha 
\vec E\times \vec B\right)\right] \label{eq:conservadas3}.
\end{eqnarray}
As it can be seen, the energy is modified but the momentum and angular 
momentum remain unchanged.\\

\noindent For the general case it is valid that
\begin{eqnarray}
& &\frac{1}{c}\frac{d {\cal E}}{dt}=-\int d\vec x \vec E\cdot \vec J
-\frac{c}{4\pi} \oint da \left(\vec E\times f\left(\nabla^{2}\right)
\vec B\right)\cdot \hat n \nonumber\\
& & {}+\frac{1}{8\pi} \int d\vec x \left( \vec B\cdot f\left(\nabla^{2}\right)
\frac{\partial \vec B}{\partial t} -\frac{\partial \vec B}{\partial t}
\cdot f\left(\nabla^{2}\right) \vec B\right).
\end{eqnarray}
The last integral in this expression is a boundary term. In fact, carrying out
a perturbative expansion we find
\begin{eqnarray}
& &\frac{1}{8\pi} \int d\vec x \left( \vec B\cdot f\left(\nabla^{2}\right)
\frac{\partial \vec B}{\partial t} -\frac{\partial \vec B}{\partial t}
\cdot f\left(\nabla^{2}\right) \vec B\right) =-\frac{c}{4\pi} \oint da u_{l}n_{l},
\end{eqnarray}
with
\begin{eqnarray}
u_{l}&=&\frac{1}{2c} \Big[ a_{2}\big(\partial_{t}B_{i}\partial_{l}B_{i}
-B_{i}\partial_{l}\partial_{t}B_{i}\big)
+a_{3}\big(\partial_{t}B_{i}\nabla^{2}\partial_{l}B_{i}-B_{i}\nabla^{2}
\partial_{t}\partial_{l}B_{i}\nonumber\\
&+&\partial_{m}B_{i}\partial_{m}\partial_{l}\partial_{t}B_{i}-\partial_{t}
\partial_{m}B_{i}\partial_{m}\partial_{m}B_{i}\big) +\cdots \Big].\quad
\end{eqnarray}
Hence, 
\begin{equation}
\frac{1}{c}\frac{d{\cal E}}{dt}=-\int d\vec x\vec E\cdot\vec J-\frac{c}{4\pi}
\oint da\left(\vec E\times f\left(\nabla^{2}\right)\vec B+\vec u\right)
\cdot\hat n.
\end{equation}
Notice that in this case the energy flux is not directly related to the
momentum $\vec P$.\\

\noindent One also finds that
\begin{eqnarray}
& &\frac{dP_{i}}{dt}=-\int d\vec x\left(\alpha\rho\vec E+\frac{\vec J\times
\vec B}{c}\right)_{\!\! i}+\frac{1}{4\pi}\oint da \tilde\tau_{ij}n_{j}
\nonumber\\
& &{}+\frac{1}{8\pi}\int d\vec x \left[\partial_{i}\vec B\cdot 
f\left(\nabla^{2}\right)\vec B- \vec B\cdot f\left(\nabla^{2}\right)
\partial_{i}\vec B\right],\qquad
\label{eq:uam-esta}
\end{eqnarray}
with
\begin{eqnarray}
\tilde \tau_{ij}&=& \alpha E_{i}E_{j} +B_{j}f\left(\nabla^{2}\right)B_{i}
-\frac{\delta_{ij}}{2}\left(\alpha \vec E\cdot \vec E +\vec B \cdot 
f\left(\nabla^{2}\right)\vec B \right).
\end{eqnarray}
The last term in (\ref{eq:uam-esta}) is a boundary integral as
\begin{eqnarray}
& & \frac{1}{8\pi c}\int d\vec x \left[\partial_{i}\vec B\cdot 
f\left(\nabla^{2}\right)\vec B-\vec B\cdot f\left(\nabla^{2}\right)
\partial_{i}\vec B\right]\nonumber\\
& &=\int d\vec x \partial_{j}\Big[ a_{2}\Big(
\partial_{i}\vec B\cdot\partial_{j}\vec B-\big(\partial_{j}\partial_{i}
\vec B\big)\cdot \vec B\Big)\nonumber\\
& & +a_{3}\Big(\partial_{i}B_{l}\nabla^{2}\partial_{j}B_{l}
-\big(\nabla^{2}\partial_{i}\partial_{j}B_{l}\big)B_{l}\nonumber\\ 
& & {}+\big(\partial_{j}\partial_{m}\partial_{i}B_{l}\big)\partial_{m}B_{l}
-\big(\partial_{i}\partial_{m}B_{l}\big)\big(\partial_{j}\partial_{m} B_{l}
\big)\Big)+\cdots \Big].\nonumber\\
\end{eqnarray}
Therefore,
\begin{equation}
\frac{dP_{i}}{dt}=-\int d\vec x \left(\alpha \rho \vec E+\frac{\vec J
\times \vec B}{c}\right)_{\!\! i}+\frac{1}{4\pi } \oint da \tau_{ij}n_{j},
\end{equation}
where the stress tensor $\tau_{ij}$ is given by
\begin{eqnarray}
& &\tau_{ij}=\alpha E_{i}E_{j} +B_{j}f\left(\nabla^{2}\right)B_{i}
-\frac{\delta_{ij}}{2}\Big(\alpha \vec E\cdot \vec E +\vec B 
f\left(\nabla^{2}\right)\vec B \Big)\nonumber\\
& &{}+a_{2}\Big(
\partial_{i}\vec B\cdot\partial_{j}\vec B-\big(\partial_{j}\partial_{i}
\vec B\big)\cdot \vec B\Big)+a_{3}\Big(\partial_{i}B_{l}\nabla^{2}\partial_{j}B_{l}
-\big(\nabla^{2}\partial_{i}\partial_{j}B_{l}\big)B_{l}\nonumber\\ 
& &{}+\big(\partial_{j}\partial_{m}\partial_{i}B_{l}\big)\partial_{m}B_{l}
-\big(\partial_{i}\partial_{m}B_{l}\big)\big(\partial_{j}\partial_{m} B_{l}
\big)\Big)+\cdots\qquad
\end{eqnarray}
For $\rho(\vec x)=e\delta^{3}\left (\vec x-\vec x^{\prime}\right)$ and
$\vec J(\vec x)=e \vec v \delta^{3}\left (\vec x-\vec x^{\prime}\right)$
one obtains
\begin{equation}
\int d\vec x \left(\alpha \rho \vec E +\frac{1}{c}\vec J\times 
\vec B\right)_{\! i}=e\alpha \vec E+\frac{e}{c} \vec v\times \vec B,
\end{equation} 
which is a modified Lorentz force. Note that when $\vec B=0$ this yields a
modified electric force of the form $e^{\prime}\vec E$ with $e^{\prime}=\alpha e$.
That is, the electric charge gets modified to $e^{\prime}=\alpha e$ and so, for 
instance, the electric potential of a point charge becomes $\phi=\alpha e/r$. 
A similar modification occurs to the gravitational potential of a particle 
\cite{blas:gnus}.\\

\subsection{ The scale transformations}

Note that equations (\ref{eq:escala-ecuaciones1})-(\ref{eq:escala-ecuaciones2}) are invariant under the scale transformations
\begin{equation}
\vec E \to b^{-(z+1)} \vec E, \quad\! \vec B \to b^{-2} \vec B,\quad\! 
t\to b^{z}t,\quad\! \vec x \to b \vec x.\label{eq:escalazgeneral}
\end{equation}
Under these, the action transforms as
\begin{eqnarray}
S\to b^{(1-z)}S.
\end{eqnarray}
Therefore, only if $z=1$ the transformations (\ref{eq:escalazgeneral})
are symmetries of the action. In this case, by Noether's theorem, we have
the conserved quantity
\begin{equation}
D_{z=1}=\frac{1}{4\pi c}\int d\vec x \left(\alpha \vec E\times \vec B\right)
\cdot \vec x-\frac{t}{c} {\cal E},
\end{equation}
which is the generator of dilations. For arbitrary $z$ the transformations (\ref{eq:escalazgeneral}) 
are not symmetries of the action, but it can be shown that the quantity
\begin{equation}
D=\frac{1}{4\pi c}\int d\vec x \left(\alpha \vec E\times \vec B\right)
\cdot \vec x-\frac{t}{c} {\cal E}+\int dt U,
\end{equation}
where 
\begin{equation}
U=\int d\vec x \left( a_{2}\vec B\cdot\nabla^{2}\vec B+2a_{3}\vec B\cdot 
\left(\nabla^{2}\right)^{2} \vec B+\cdots \right),
\end{equation}
is conserved. Thus  this quantity is related to the scale transformations.

\section{Duality transformations and the question of the 
magnetic monopole}
\label{sn:dual}

The usual Maxwell's equations in vacuum are invariant under the duality 
transformations
\begin{equation}
\big(\vec E,\vec B\big) \, \to \, \big(\!-\vec B,\vec E\big).
\label{eq:dual}
\end{equation} 
Introducing these into the modified Maxwell's equations (\ref{eq:m-general-1})-(\ref{eq:m-general-4})
one finds the dual equations
\begin{eqnarray}
\vec\nabla\cdot \vec E&=&0,\qquad 
\vec\nabla\times \left( f\left(\nabla^{2} \right) \vec E\right)=
-\frac{\alpha}{c}\frac{\partial \vec B}{\partial t},\label{eq:dual-l-2}\\
\vec \nabla\cdot \vec B&=&0,\qquad
\vec\nabla\times \vec B =\frac{1}{c}\frac{\partial \vec E}{\partial t}.
\label{eq:dual-l-4}
\end{eqnarray}
Therefore, unlike the usual case, the equations are not
self-dual. Now one gets a system of modified Maxwell's equations with the 
Faraday's law modified. It can be observed that these 
equations imply the wave equations (\ref{eq:onda-1}) and (\ref{eq:onda-2}),
and so Eqs. (\ref{eq:dual-l-2})-(\ref{eq:dual-l-4}) are also consistent 
with the dispersion relation (\ref{eq:dispersion-general}).\\

Now, notice that in terms of the scalar magnetic potential $\tilde \phi$
and the electric vector potential $\tilde A_{i}$ one has
\begin{eqnarray}
B_{i}=-\left(\partial_{ct}\tilde A_{i}+\partial_{i}\tilde \phi\right),\quad 
E_{i}= \epsilon_{ijk}\partial_{j}\tilde A_{k}.
\end{eqnarray}
Dual equations (\ref{eq:dual-l-2})-(\ref{eq:dual-l-4}) can be obtained
from the dual action
\begin{eqnarray}
S&=&\int cdt d\vec x {\cal  L}= \int cdt d\vec x \left(\alpha B_{i}B_{i} 
- E_{i}f\left(\nabla ^{2}\right) E_{i}\right)\nonumber\\
&=&\int cdt d\vec x \left(\alpha  B_{i} B_{i} -\frac{1}{2} \tilde F_{ij} 
f\left(\nabla ^{2}\right)\tilde F_{ij}\right),
\end{eqnarray}
which is compatible with the dispersion relation 
(\ref{eq:dispersion-general}) and in particular with (\ref{eq:Horava}).\\

Let us now consider the modified Maxwell's equations including magnetic
monopoles
\begin{eqnarray}
\vec\nabla\cdot \vec E&=&4\pi \rho_{e}, \qquad 
\vec\nabla\times \vec E=-\left( \frac{4\pi}{c}\vec J_{m}
+\frac{1}{c}\frac{\partial \vec B}{\partial t}\right)\label{eq:m-m2},\\
\vec \nabla\cdot \vec B&=& 4\pi \rho_{m},\qquad
\vec\nabla\times\left(f\left(\nabla^{2}\right)\vec B\right)=
\frac{4\pi}{c}\vec J_{e}+\frac{\alpha}{c}\frac{\partial\vec E}{\partial t}.
\label{eq:m-m4}
\end{eqnarray}
By performing the duality transformations
\begin{eqnarray}
\big(\vec E,\vec B\big) \to \big(-\vec B,\vec E\big),\quad 
\big(\rho_{e},\rho_{m}\big) \to \big(-\rho_{m},\rho_{e}\big),\quad 
\big(\vec J_{e},\vec J_{m}\big) \to \big(-\vec J_{m},\vec J_{e}\big),
\label{eq:transformacion-dualidad}
\end{eqnarray} 
one obtains the dual equations
\begin{eqnarray}
\vec\nabla\cdot \vec E&=&4\pi \rho_{e}, \qquad 
\vec\nabla\times \left( f\left(\nabla^{2}\right)\vec E\right)=-\left(
\frac{4\pi}{c}\vec J_{m}+\frac{\alpha}{c}\frac{\partial\vec B}{\partial t}
\right)\label{eq:m-m2-b},\\
\vec \nabla\cdot \vec B&=& 4\pi \rho_{m},\qquad
\vec\nabla\times \vec B=\frac{4\pi}{c}\vec J_{e}+
\frac{1}{c}\frac{\partial \vec E}{\partial t}.\label{eq:m-m4-b}
\end{eqnarray}
Then, if we have solutions of the equations (\ref{eq:m-m2})-(\ref{eq:m-m4}), 
by duality transformations (\ref{eq:transformacion-dualidad}), solutions to (\ref{eq:m-m2-b})-(\ref{eq:m-m4-b}) are obtained.\\
 
Some topics on the magnetic monopole for the case of $z=2$
have been discussed in \cite{monopolo:gnus}.

\section{Green Function}

Introducing the definitions of the potentials into (\ref{eq:m1}) 
and (\ref{eq:m4}) we find
\begin{eqnarray}
\nabla^{2}\phi +\frac{1}{c}\frac{\partial\vec\nabla\cdot\vec A}{\partial t}
&=&-4\pi \rho,\\
f\left(\nabla^{2}\right) \nabla^{2}\vec A-\frac{\alpha}{c^{2}} 
\frac{\partial^{2}\vec  A}{\partial t^{2}}
&-&\vec \nabla \cdot\left(  f\left(\nabla^{2}\right)\vec\nabla\cdot\vec A
+\frac{\alpha}{c}\frac{\partial \phi}{\partial t}\right)=-\frac{4\pi}{c} \vec J.
\end{eqnarray}
 Using the Coulomb gauge, $\vec \nabla\cdot \vec A=0,$ in the static case we have 
\begin{eqnarray}
\nabla^{2}\phi =-4\pi \rho,\quad 
f\left(\nabla^{2}\right) \nabla^{2}\vec A=-\frac{4\pi}{c} \vec J.
\end{eqnarray}
If $f\left(\nabla^{2}\right)=a_{z} \left(\nabla^{2}\right)^{z-1}$ we obtain
\begin{eqnarray}
\nabla^{2}\phi =-4\pi \rho,\qquad
\left(\nabla^{2}\right)^{z}\vec A=-\frac{4\pi}{ca_{z}} \vec J.
\end{eqnarray}
The solution for  $\vec A$ is given by 
\begin{eqnarray}
\vec A(\vec x)=-\frac{4\pi}{ca_{z}}\int d\vec x_{1} G\left(\vec x,\vec x_{1}\right) \vec J\left(\vec x_{1}\right),
\end{eqnarray}
where
\begin{eqnarray}
G\left(\vec x,\vec x_{1}\right)=\int d\vec x_{z} d\vec x_{z-1}\cdots d\vec x_{2} G_{0}\left(\vec x,\vec x_{z}\right) G_{0}\left(\vec x_{z},\vec x_{z-1}\right) \cdots   G_{0}\left(\vec x_{2},\vec x_{1}\right),
\label{eq:green}
\end{eqnarray}
and
\begin{eqnarray}
G_{0}\left(\vec x_{a},\vec x_{b}\right)=\frac{-1}{4\pi |\vec x_{a}-\vec x_{b}|},
\end{eqnarray}
is the Green function for the Maxwell static theory. Remarkably the Green function (\ref{eq:green}) seems like
a quantum correction of order $z$ to the Maxwell's static two-point function \cite{peskin-1:gnus,peskin-2:gnus}.\\

As it may be seen, the Lorentz gauge condition is substituted by
\begin{equation}
f\left(\nabla^{2}\right) \vec \nabla\cdot \vec A+\frac{\alpha}{c}
\frac{\partial \phi}{\partial t}=0.
\end{equation}
In terms of this, one obtains
\begin{eqnarray}
\left( f\left(\nabla^{2}\right) \nabla^{2} -\frac{\alpha}{c^{2}} 
\frac{\partial^{2}}{\partial t^{2}}\right)\phi
&=&-4\pi f\left(\nabla^{2}\right)\rho,\\
\left(  f\left(\nabla^{2}\right) \nabla^{2} -\frac{\alpha}{c^{2}}
\frac{\partial^{2}}{\partial t^{2}}\right)\vec A
&=&-\frac{4\pi}{c} \vec J.
\end{eqnarray}
For this case the potentials are
\begin{eqnarray}
\phi(\vec x,t)&=&\int d\vec x^{\prime}dt^{\prime} 
\bigg[f\left( \vec \nabla^{\prime 2}\right)\rho (\vec x^{\prime},t^{\prime})\bigg] 
g\left (\vec R, \tau \right),\nonumber \\
\vec A(\vec x,t)&=&\int d\vec x^{\prime}dt^{\prime} 
\vec J(\vec x^{\prime}) g\left (\vec R, \tau \right), \quad \vec R=\vec x-\vec x^{\prime}, \tau=t-t^{\prime}\nonumber
\end{eqnarray}
where
\begin{eqnarray}
g\left (\vec R, \tau \right)=-\frac{1}{4\pi^{3}}
\int d\vec kd\omega e^{ i\left( \vec k\cdot\vec R-\omega \tau \right)} \tilde g\left(k^{2},\omega \right)
\end{eqnarray}
with 
\begin{eqnarray}
 \tilde g\left(k^{2},\omega \right)=\frac{c^{\prime 2} }{ \omega^{2} -k^{2}c^{\prime 2}f\left(-k^{2}\right) }
\end{eqnarray}
If $k^{2}c^{\prime 2}f\left(-k^{2}\right)=k^{2}c^{\prime 2}+ a_{z} \left(k^{2}\right)^{z},$ we can find
\begin{eqnarray}
\tilde g\left(k^{2},\omega \right)=\frac{c^{\prime 2} }{ \omega^{2} -k^{2}c^{\prime 2}- a_{z} \left(k^{2}\right)^{z}}.
\label{eq:propagador}
\end{eqnarray}
At low energies, the propagator is dominated by the Maxwell theory 
\begin{eqnarray}
\tilde g\left(k^{2},\omega \right)= c^{\prime 2}\Bigg( \frac{1}{ \omega^{2} -k^{2}c^{\prime 2} }+\frac{1}{ \omega^{2} -k^{2}c^{\prime 2} }a_{z} \left(k^{2}\right)^{z}\frac{1}{ \omega^{2} -k^{2}c^{\prime 2} }\
+\cdots \Bigg).\label{eq:propagador1}
\end{eqnarray}
Therefore we can see this theory as a quantum correction to the Maxwell classical theory.\\

In the high energies regime, the propagator is dominated by $\tilde g\left(k^{2},\omega \right)=c^{\prime 2}/(\omega^{2} - a_{z} \left(k^{2}\right)^{z}),$ that is
\begin{eqnarray}
\tilde g\left(k^{2},\omega \right)= c^{\prime 2}\Bigg( \frac{1}{ \omega^{2} -a_{z} \left(k^{2}\right)^{z} } +\frac{1}{ \omega^{2} -a_{z} \left(k^{2}\right)^{z} } 
 k^{2}c^{\prime 2} \frac{1}{ \omega^{2} -a_{z} \left(k^{2}\right)^{z} } +\cdots \Bigg).
\label{eq:propagador2}
\end{eqnarray}
The same occurs in the Ho\v{r}ava  gravity \cite{horava:gnus}. \\

\section{Summary}
\label{sn:summ}

In this work we studied an electrodynamics consistent with anisotropic
transformations of the space-time with an arbitrary dynamic exponent $z$.
The equations of motion and conserved quantities were obtained.  It was shown that the propagator of this theory  can be regarded as a quantum correction to  the usual Maxwell's propagator. Also, it was 
shown that the momentum and angular momentum remain unchanged, but their 
conservation laws have modifications. It was shown that both the speed of light and electric charge run  
with $z$. The existence of an
additional conserved quantity that changes with $z$ was also shown which, if $z=1,$ is reduced the generator of dilations.
In addition, the question of the magnetic monopole was considered, and this
lead to showing a dual electrodynamics invariant under the same 
anisotropic transformations. \\

In a further work we  will study the quantum version of this model. 
From  (\ref{eq:propagador2}) a better UV behavior than the usual case is expected.
Moreover from propagator (\ref{eq:propagador}) and equations (\ref{eq:propagador1})-(\ref{eq:propagador2})
we can expect that this model has two fixed points.
We hope to prove this affirmation with the renormalization group and  study the crossover
between these fixed points. We will also couple this field with matter.


\begin{thebibliography}{99}

\bibitem{condensada:gnus}
P.~C.~Hohenberg, B.~I.~Halperin, {\it Theory of Dynamic Critical 
Phenomena,} Rev. Mod. Phys. {\bf 49}, 435 (1977).

\bibitem{cardy:gnus}
J.~L.~Cardy, {\it Scaling and Renormalization in Statistical Physics}
(Cambridge, UK, 1996).

\bibitem{milgrom:gnus}
M.~Milgrom, {\it Nonlinear Conformally Invariant Generalization of the 
Poisson Equation to $D>2$ Dimensions,} Phys. Rev. E {\bf 56}, 1148 (1997).

\bibitem{mond-horava:gnus}
J.~M.~Romero, R. Bernal-Jaquez, O. Gonzalez-Gaxiola, {\it Is 
Possible to Relate MOND with Ho\v{r}ava Gravity?} In press (to appear in 
Mod. Phys. Lett. A), e-Print: arXiv:1003.0684 [hep-th].

\bibitem{string-1:gnus}
D.~T.~Son, {\it Toward an AdS/Cold Atoms Correspondence: A Geometric 
Realization of the Schroedinger Symmetry,} Phys. Rev. D {\bf 78}, 046003 
(2008).

\bibitem{string-2:gnus}
K.~Balasubramanian, J.~McGreevy, {\it Gravity Duals for 
Non-relativistic CFTs}, Phys. Rev. Lett. {\bf 101}, 061601 (2008).

\bibitem{herzog1:gnus}
S.~Gubser, C.~Herzog, S.~Pufu, T.~Tesileanu, {\it Superconductors from 
Superstrings,} Phys. Rev. Lett. {\bf 103}, 141601 (2009).

\bibitem{herzog2:gnus}
C.~Herzog, {\it  Lectures on Holographic Superfluidity and 
Superconductivity,} J. Phys. A {\bf 42}, 343001 (2009).

\bibitem{horava:gnus}
P.~Ho\v{r}ava, {\it Quantum Gravity at a Lifshitz Point,} 
Phys. Rev. D {\bf 79}, 084008 (2009).

\bibitem{brandenberger:gnus}
R.~Brandenberger, {\it Matter Bounce in Ho\v{r}ava-Lifshitz Cosmology,} 
Phys. Rev. D {\bf 80}, 043516 (2009).

\bibitem{mukohyama:gnus}
S.~Mukohyama, {\it Dark Matter as Integration Constant in 
Ho\v{r}ava-Lifshitz Gravity,} Phys. Rev. D {\bf 80}, 064005 (2009).
 
\bibitem{bibhas-1:gnus}
B.~R.~Majhi, {\it Hawking radiation and black hole spectroscopy in 
Horava-Lifshitz gravity,} Phys. Lett. B {\bf 686}, 49 (2010).

\bibitem{bibhas-2:gnus}
E.~O.~Colgain, H.~Yavartanoo, {\it  Dyonic solution of Horava-Lifshitz 
Gravity,} JHEP {\bf 0908}, 021 (2009).

\bibitem{romero1:gnus}
J. M.~Romero, V.~Cuesta, J.~A.~Garcia, J.~D.~Vergara, 
{\it Conformal Anisotropic Mechanics and the Ho\v{r}ava Dispersion 
Relation,} Phys. Rev. D {\bf 81}, 065013 (2010).

\bibitem{romero2:gnus}
T.~Suyama, {\it Notes on Matter in Horava-Lifshitz Gravity,}
JHEP {\bf 01}, 093 (2010).

\bibitem{romero3:gnus}
D.~Capasso,  A.~P.~Polychronakos, {\it Particle Kinematics in 
Horava-Lifshitz Gravity,} JHEP {\bf 02}, 068 (2010).

\bibitem{romero4:gnus}
J.~Kluson, {\it String in Horava-Lifshitz Gravity,}
e-Print: arXiv:1002.2849 [hep-th].

\bibitem{romero5:gnus}
M.~Eune, W.~Kim, {\it Note on an action for a particle in the 
Ho\v{r}ava-Lifshitz Gravity,} e-Print: arXiv:1003.4052 [hep-th].

\bibitem{romero6:gnus}
A.~E.~Mosaffa, {\it On Geodesic Motion in Horava-Lifshitz Gravity,}
e-Print: arXiv:1001.0490 [hep-th]. 


\bibitem{pag-1:gnus}
M.~Li, Y.~Pang, {\it A Trouble with Ho\v{r}ava-Lifshitz Theory,}
JHEP {\bf 0908}, 015 (2009).

\bibitem{pag-2:gnus}
D.~Blas, O.~Pujolas, S.~Sibiryakov, {\it On the Extra Mode and 
Inconsistency of Ho\v{r}ava Gravity,} JHEP {\bf 0910}, 029 (2009).

\bibitem{pag-3:gnus}
C.~Charmousis, G.~Niz, A.~Padilla,  P.~M.~Saffin,
{\it Strong coupling in Horava gravity,} JHEP {\bf 0908}, 070 (2009).

\bibitem{pag-4:gnus}
M.~Henneaux, A.~Kleinschmidt, G.~L.~G\'omez, {\it A Dynamical 
Inconsistency of Ho\v{r}ava Gravity,} Phys. Rev. D {\bf 81}, 064002 
(2010).

\bibitem{pag-5:gnus}
C.~Bogdanos, E.~N.~Saridakis, {\it Perturbative Instabilities in 
Ho\v{r}ava Gravity,} Class. Quant. Grav. {\bf 27}, 075005 (2010).

\bibitem{pag-6:gnus}
A. Kobakhidze, {\it On the infrared limit of Horava's gravity with 
the global Hamiltonian constraint,} e-Print: arXiv:0906.5401 [hep-th]. 


\bibitem{venezuela:gnus}
J.~Bellorin, A.~Restuccia, {\it On the consistency of the Horava Theory,}
e-Print: arXiv:1004.0055 [hep-th]. 

\bibitem{blas:gnus}
D.~Blas, O.~Pujolas,  S.~Sibiryakov, {\it Consistent Extension of 
Ho\v{r}ava Gravity,} Phys. Rev. Lett. {\bf 104}, 181302 (2010).


\bibitem{blas-1:gnus}
P. Horava,  C. M. Melby-Thompson, {\it General Covariance in Quantum Gravity at a Lifshitz Point,}
e-Print: arXiv:1007.2410 [hep-th] 

\bibitem{blas-2:gnus}
D. Blas, O. Pujolas, S. Sibiryakov, 
{\it Models of non-relativistic quantum gravity: the good, the bad and the healthy,}
 e-Print: arXiv:1007.3503 [hep-th] 

\bibitem{tony:gnus}
I.~Kimpton,  A.~Padilla, {\it Lessons from the decoupling limit of 
Horava gravity,} JHEP {\bf 1007} 014 (2010),  e-Print: arXiv:1003.5666 [hep-th].

\bibitem{das:gnus}
S.~R.~Das, Ganpathy Murthy, {\it Compact $z=2$ Electrodynamics in $2+1$ 
Dimensions: Confinement with Gapless Modes,} Phys. Rev. Lett. {\bf 104}, 
181601 (2010).

\bibitem{andreev:gnus}
O.~Andreev, {\it Generating Functional for Gauge Invariant Actions: 
Examples of Nonrelativistic Gauge Theories,}
Int. J. Mod. Phys. A {\bf 25}, 2087 (2010).

\bibitem{nakayama:gnus}
Y.~Nakayama, {\it Superfield Formulation for Non-Relativistic 
Chern-Simons-Matter Theory,} Lett. Math. Phys. {\bf 89}, 67 (2009).

\bibitem{pavlopoulos:gnus}
T. G. Pavlopoulos, {\it Breakdown of Lorentz Invariance,} 
Phys. Rev. {\bf 159} 1106 (1967). 


\bibitem{tpsalis-1:gnus}
J.~Alexandre, K.~Farakos,  A.~Tsapalis,
{\it Liouville-Lifshitz theory in 3+1 dimensions,}
Phys. Rev. D {\bf 81}, 105029 (2010).


\bibitem{tpsalis-2:gnus}
J.~Alexandre, K.~Farakos, P.~Pasipoularides,  A.~Tsapalis, 
{\it Schwinger-Dyson approach for a Lifshitz-type Yukawa model,}
Phys. Rev. D {\bf 81}, 045002 (2010).


\bibitem{amanda:gnus}
G. Bertoldi, B. A. Burrington, Amanda Peet, 
{\it Black Holes in asymptotically Lifshitz spacetimes with arbitrary critical exponent,}
Phys. Rev. D {\bf 80} 126003 (2009).

\bibitem{hartnoll:gnus}
S. A. Hartnoll, K. Yoshida, {\it Families of IIB duals for nonrelativistic CFTs,}
JHEP {\bf 0812} 071 (2008).

\bibitem{malda:gnus}
J. Maldacena, D. Martelli, Y. Tachikawa, 
{\it Comments on string theory backgrounds with non-relativistic conformal symmetry,}
JHEP {\bf 0810} 072 (2008).

\bibitem{blau:gnus}
M. Blau, J. Hartong, B. Rollier,
{\it Geometry of Schrodinger Space-Times, Global Coordinates, and Harmonic Trapping,} 
JHEP {\bf 0907} 027 (2009).

\bibitem{fradkin:gnus}
E.~Fradkin, D.~A.~Huse, R.~Moessner, V.~Oganesyan, S.~L.~Sondhi,
{\it On bipartite Rokhsar-Kivelson points and Cantor deconfinement,}
Phys. Rev. B {\bf 69}, 224415 (2004).

\bibitem{horava3:gnus}
P.~Ho\v{r}ava, {\it Quantum Criticality and Yang-Mills Gauge Theory,}
e-Print: arXiv:0811.2217 [hep-th].

\bibitem{chen:gnus}
B.~Chen, Q.~Huang, {\it Field Theory at a Lifshitz Point,} 
Phys. Lett. B {\bf 683}, 108 (2010).

\bibitem{hussin:gnus}
V.~Hussin, M.~Jacques, {\it On Nonrelativistic Conformal Symmetries 
and Invariant Tensor Fields,} J. Phys. A {\bf 19}, 3471 (1986).

\bibitem{yifu:gnus}
Y.~Cai, M.~Li,  X.~Zhang,
{\it Testing the Lorentz and CPT Symmetry with CMB polarizations and 
a non-relativistic Maxwell Theory,} JCAP {\bf 1001}, 017 (2009).

\bibitem{nakayam:gnus}
Y.~Nakayama, {\it Superfield Formulation for Non-Relativistic 
Chern-Simons-Matter Theory,} Lett. Math. Phys. {\bf 89}, 67 (2009).

\bibitem{kiritsis:gnus}
E.~Kiritsis, G.~Kofinas, {\it Horava-Lifshitz Cosmology,}
Nucl. Phys. B {\bf 821}, 467 (2009).

\bibitem{maeda:gnus}
S.~Maeda, S.~Mukohyama, T.~Shiromizu, {\it Primordial Magnetic Field 
from Non-inflationary Cosmic Expansion in Horava-Lifshitz Gravity,}
Phys. Rev. D {\bf 80}, 123538 (2009).

\bibitem{gruss:gnus}
E.~Gruss, {\it Black Holes in Ho\v{r}ava Gravity with Higher Derivative 
Magnetic Terms,} e-Print: arXiv:1005.1353 [hep-th].

\bibitem{monopolo:gnus}
R.~Moessner, S.~L.~Sondhi, {\it  Three-dimensional 
Resonating-valence-bond Liquids and Their Excitations,} Phys. Rev. B 
{\bf 68}, 184512 (2003).


\bibitem{peskin-1:gnus}
M. E. Peskin, D. V. Schroeder, {\it An introduction to quantum field theory,}
Westview Press, United States of America (1995).

\bibitem{peskin-2:gnus}
W. Greiner, J. Reinhardt, {\it Field quantization,}
Springer-Verlag, Berlin (1996).


\end{thebibliography}
\end{document}